# Fundamentals of PV Efficiency Interpreted by a Two-Level Model

Muhammad A. Alam and M. Ryyan Khan
*Electrical and Computer Engineering Department, Purdue University*
e-mail: alam@purdue.edu

*Abstract*—Elementary physics of photovoltaic energy conversion in a two-level atomic PV is considered. We explain the conditions for which the Carnot efficiency is reached and how it can be exceeded! The loss mechanisms – thermalization, angle entropy, and below-bandgap transmission – explain the gap between Carnot efficiency and the Shockley-Queisser limit [1]. Wide varieties of techniques developed to reduce these losses (e.g., solar concentrators, solar-thermal, tandem cells, etc.) are reinterpreted by using a two level model. Remarkably, the simple model appears to capture the essence of PV operation and reproduce the key results and important insights that are known to the experts through complex derivations [2–4].

I. INTRODUCTION

Renewable energy is a topic of broad, current interest. Photovoltaic devices – which convert radiant energy from the sun into electrical energy – offer a promising source for renewable energy. Since a solar cell is essentially a p-n junction diode illuminated by sunlight, its performance can be understood in terms of classical diode equation coupled with a current source to account for photogeneration [5]. The key parameters that dictate the efficiency of energy conversion such as short circuit current, open circuit voltage, and fill factor are easily related to basic diode parameters such as doping densities, base and emitter thicknesses, bulk and interface recombinations, etc. [6][7] and this detailed understanding of device operation has led to impressive gain in PV efficiency since 1950s. Coupled with sophisticated process engineering, classical solar cells are beginning to approach the fundamental limits of energy conversion [8–10]. Future progress will depend on understanding the origin of the remaining gap between the 'fundamental' and practical limits of PV efficiency.

In this paper, we explain the fundamental limits of energy conversion when a solar cell is viewed as a 'photon engine' operating between two reservoirs, i.e. the sun and the environment. We discuss the physics of a photovoltaic operation of a collection of two-level atoms. We find that the model anticipates – transparently and intuitively – the fundamental issues of efficiency of a solar cells; historically many of these issues have been derived from far more complicated arguments [2–4]. The functional relationships derived for the two-level model correctly anticipates the corresponding results for two and three dimensional bulk solar cells, except for the numerical coefficients that depend on system dimensionality.

II. PHYSICS OF IDEALIZED 2-LEVEL SYSTEMS

*A. A 2-level system*

Consider a set of two-level 'atoms' immersed in an isotropic, three dimensional field of photons. An analogous problem arises when discussing the physics of photosynthesis in pigment molecules of marine diatoms immersed in fluid, illuminated by multiply reflected, diffuse light [11]. We will consider discrete levels, although as long as the widths of the bands are much narrower than the energy of the photons, the conclusions apply. Our goal in this section is to show that if we could connect these 'atoms' with weak probes to extract the photogenerated electrons, we might be able to achieve or even exceed the Carnot efficiency – the ultimate limit of energy conversion in any thermodynamic engine.

*B. Two-level system illuminated by a monochromatic sun*

Typically, if the atoms remain in equilibrium with its surrounding of phonons and photons, the relative populations of atoms in the ground state $E_2$ versus those in the excited states $E_1$ are simply given by the Fermi-Dirac (F-D) statistics, i.e.,

$$f_i = \frac{1}{e^{\theta_i/T_D} + 1} \qquad (1)$$

where $k_B \theta_i \equiv E_i - \mu_{i,m}$, where $E_i = E_1, E_2$ and $\mu_{i,m}$ is the chemical potential associated with the state. $(\mu_1 - \mu_2)$ is not necessarily zero.

Since F-D statistics is interpreted as the probability of occupation of a state in a bulk semiconductor, one may wonder regarding the meaning of a F-D distribution of a two-level system, where the atoms can either be in the excited state or in the ground states. Here the probability of occupation reflects the property of the ensemble, i.e. the *fraction* of atoms in the up (or down) states is characterized by the F-D distribution of those states, appropriately normalized so that the sum of the atoms in the two levels equals the total number of atoms $N$.

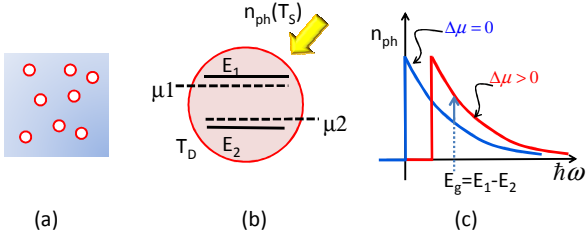

**Figure 1: (a) A collection of 2-level atoms. (b) A 2-level energy system illuminated by photons. (c) The Bose-Einstein distribution.**

The external isotropic monochromatic illumination of these atoms (see Fig. 1) changes the relative population by rebalancing the absorption and emission rates. The absorption or the 'up' transition is given by

$$U(E_2 \to E_1) = A f_2 (1 - f_1) n_{ph}, \quad (2)$$

while the emission rate or 'down' transition is given by

$$D(E_1 \to E_2) = A f_1 (1 - f_2)(n_{ph} + 1). \quad (3)$$

Here, $A$ is a constant, the extra 1 on the right hand side transition describes the spontaneous emission (see Feynman, vol. 3, Chapter. 4 for more detailed discussion), and $n_{ph}$ is the Bose Einstein distribution for isotropic photons, given by

$$n_{ph}(T_s) = \frac{1}{e^{[(E_1 - E_2) - (\mu_1 - \mu_2)_s]/k_B T_s} - 1}, \quad (4)$$

see Fig. 1c. Here, the subscript S is a reminder that we are talking about photons coming from a 'monochromatic sun'. The form of Eq. (4) may be unfamiliar but easily derived: Assume the sun to be an isolated box of atoms and photons in equilibrium at temperature $T_S$, equate (2) and (3), and then substitute (1) for F-D statistics, and solve for $n_{ph}(T_S)$. Although the sun is powered by internal nuclear reaction, measurement of the solar spectrum shows that $(\mu_1 - \mu_2)_S \equiv \Delta \mu_s \approx 0$ [5][12]. We will use this assumption for the following discussion.

Under 'open-circuit condition' for the two level system kept at temperature $T_D$ (with distribution defined by Eq. (1)) illuminated by photons from a source at temperature $T_S$ (with distribution defined by Eq. (4)), the absorption must be balanced by emission, i.e., $U = D$,

$$f_1 (1 - f_2)(n_{ph} + 1) = f_2 (1 - f_1) n_{ph}. \quad (5)$$

Inserting Eq. (1) and (4) in Eq. (5), we find – after a few lines of algebra – that

$$\frac{E_2 - \mu_2}{T_D} + \frac{E_2 - E_1}{T_S} = \frac{E_1 - \mu_1}{T_D}. \quad (6)$$

Or, equivalently,

$$V_{oc} \equiv (\mu_1 - \mu_2)_D = (E_1 - E_2)\left[1 - \frac{T_D}{T_S}\right]. \quad (7)$$

We should notice the appearance of the Carnot factor involving the ratio of the 'device' temperature and the temperature of the sun.

Now if we could attach a pair of weak probes to each of the atoms and if the photon flux R from the sun is small, then the energy input to the ensemble of atoms is $(E_1 - E_2) \times R \times N$, while the maximum energy output is $(\mu_1 - \mu_2)_D \times R \times N$, so that the efficiency $\eta$ is given by

$$\eta = \frac{(\mu_1 - \mu_2)_D R \times N}{(E_1 - E_2) R \times N} = \left[1 - \frac{T_D}{T_S}\right]. \quad (8)$$

In this limit, a photon engine is just another form of 'heat' engine connected between two reservoirs of temperature $T_S$ and $T_D$, described by the Carnot formula. Assuming that the atoms are at room temperature ($T_D = 300K$) and the sun is a blackbody with $T_S = 6000K$, the efficiency is

$$\eta = \left[1 - \frac{300}{6000}\right] = 0.95.$$

The conventional PV conversion efficiency limit is 33%—the so called Shockley-Queisser limit [1]. This dramatic difference of the efficiency between the two-level atomic PV (Eq. (8)) and that of the practical 3D solar cells lies in three factors: the sun is far away and occupies (as a disk) a small fraction of the sky, the dimensionality of the solar cell, and the peculiar definition of Shockley-Queisser efficiency. The meaning of the preceding sentence will become clear in the later sections of the paper.

Just to complete the equivalence of the photon engine with typical reversible thermodynamic engine, let us calculate (based on Fig. 2) the entropy produced in the conversion process by summing up over all the processes involved, i.e.,

$$S = \sum \frac{Q}{T} = \frac{Q_2}{T_D} + \frac{Q_{ph}}{T_S} - \frac{Q_1}{T_D}$$

$$= \frac{E_2 - \mu_2}{T_D} + \frac{E_2 - E_1}{T_S} - \frac{E_1 - \mu_1}{T_D}$$

$$= 0$$

where $Q_1/T_D$ and $Q_2/T_D$ are the entropy generated when an electron and a hole exit the contacts respectively, and $Q_{ph}/T_S$ is the entropy produced by photogeneration (see Figure 2). This result in not surprising –since Carnot cycle is reversible, there is no net entropy production in the system. The schematic in Fig. 2b makes the analogy between a solar cell and photon engine explicit.

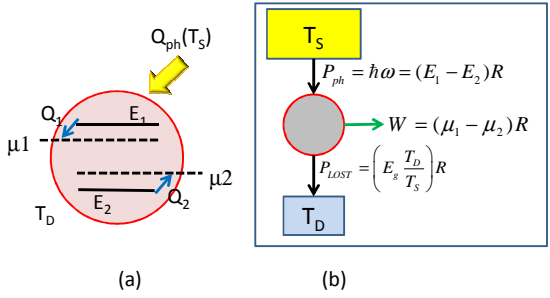

**Figure 2:** (a) The energy band for the 2-level system. (b) The energy flux balance of a 'photon engine'.

### C. Two-level atoms with multiple gaps

Let us return to our original discussion of two level atoms illuminated by diffused sunlight, not by a LED. In Sec. A, all the atoms had identical energy gaps and could absorb only at a single energy; the system achieved Carnot efficiency. If we generalize the problem, so that the ensemble includes $N_1$ atoms with $E_{G,1} \equiv (E_1^{(1)} - E_2^{(1)})$, $N_2$ atoms of $E_{G,2} \equiv (E_1^{(2)} - E_2^{(2)})$, etc. can the ensemble as a whole still achieve the Carnot efficiency?

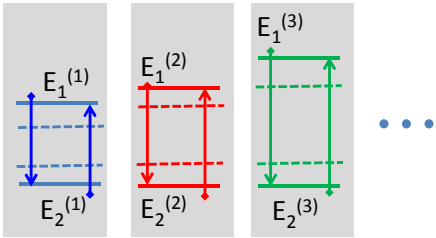

**Figure 3:** (a) Ensemble of 'non-interacting' 2-level systems having different energy gaps.

The total energy input to the system is $P_{in} = \sum_{i=1}^{M}(E_1^{(i)} - E_2^{(i)})N_i$, while the total power-output $P_{out} = \sum_{i=1}^{M}(\mu_1^{(i)} - \mu_2^{(i)})N_i$. The principle of detailed balance requires that each group of atoms is in equilibrium with the corresponding set of incident photons, i.e., $(\mu_1^{(i)} - \mu_2^{(i)}) = \eta_i(E_1^{(i)} - E_2^{(i)})$. Of course, each 2-level system operates at the Carnot efficiency ($\eta_i = \eta_1$). Taken together,

$$\eta_S = \frac{N_1(\mu_1^{(1)} - \mu_2^{(1)}) + N_2(\mu_1^{(2)} - \mu_2^{(2)}) + \ldots}{N_1(E_1^{(1)} - E_2^{(1)}) + N_2(E_1^{(2)} - E_2^{(2)}) + \ldots}$$
$$= \frac{\eta_1 N_1(E_1^{(1)} - E_2^{(1)}) + \eta_1 N_2(E_1^{(2)} - E_2^{(2)}) + \ldots}{N_1(E_1^{(1)} - E_2^{(1)}) + N_2(E_1^{(2)} - E_2^{(2)}) + \ldots}$$
$$= \eta_1.$$

The ensemble of atoms absorbing at different frequencies can still achieve Carnot efficiency – provided the atoms are isolated and energy is independently collected by weakly coupled probes attached to these 'atoms'. In the PV literature, solar cells based on such 'spectral splitting technique' has been discussed in the context of very high efficiency cells [13].

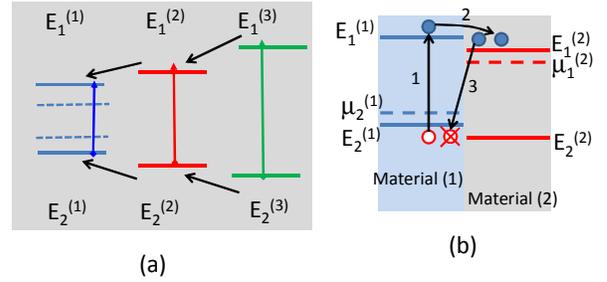

**Figure 4:** (a) Ensemble of 'interacting' 2-level systems having different energy gaps. (b) Operation of excitonic solar cell.

### III. PHYSICS OF SOMEWHAT REAL PV: ENSEMBLE OF 2-LEVEL SYSTEM

Practical limits of solar cells are well-known to be far lower than the Carnot limit. Where does the energy go? We will discuss three sources of energy loss: thermalization loss, irreversible generation of entropy due to angle mismatch, and the transmission loss – all in the context of the two-level system—to understand gap between real and the Carnot-efficient solar cells.

### A. Thermalization Loss

Let us return to the discussion in Sec. II.D, where we considered an ensemble of *independent* atoms illuminated by diffused sunlight. Instead, now we assume that the atoms are coupled – as in a solid – so that electrons can transfer from one atom to the next, see Fig. 4(a). The transfer of electrons from atoms with larger gap to atoms with smaller gap is accompanied by emission of phonons to the environment. We will assume that all the atoms can absorb photons, photon emission is only possible for atoms with the smallest energy gap, $(E_1^{(1)} - E_2^{(1)})$, and energy is only extracted at the smallest gap. In this case,

$$\eta_S = \frac{(N_1 + N_2 + N_3 + \cdots)(\mu_1^{(1)} - \mu_2^{(1)})}{N_1(E_1^{(1)} - E_2^{(1)}) + N_2(E_1^{(2)} - E_2^{(2)}) + \ldots} \quad (9)$$

$$= \eta_1 \frac{(N_1 + N_2 + N_3 + \cdots)(E_1^{(1)} - E_2^{(1)})}{N_1(E_1^{(1)} - E_2^{(1)}) + N_2(E_1^{(2)} - E_2^{(2)}) + \ldots} < \eta_1$$

The loss of efficiency is expected, as the energy absorbed in atoms with larger bandgap has been lost to thermalization (phonon emission). Moreover, one can show – by repeating the steps in Sec. A – that the process generates entropy and the system is no longer reversible. In PV parlance, this is the called the thermalization loss; it arises from coupling among atoms.

*Thermalization loss in excitonic PV*

As the simplest example of thermalization loss in coupled two-level atoms, consider an excitonic PV with a donor and an acceptor atom ((1) and (2)) linked together as a common unit, as shown in Fig. 4b. Examples of such atoms donor and acceptor atoms include P3HT and PCBM, respectively *[14], [15]*. Photons are absorbed in atom (1), generating a tightly bound electron-hole pair called an exciton (process '1'), the exciton dissociates at the (1)-(2) boundary into free electron and hole, and the electron transfers to $E_1^{(2)}$ of material (2) (process '2'). The free electron at $E_1^{(2)}$ and hole at $E_2^{(1)}$ recombine at the cross gap, giving away photons of energy $(E_1^{(2)} - E_2^{(1)})$, process '3'. The up and down transitions are given by

$$U(E_2^{(1)} \to E_1^{(1)}) = A f_2 (1 - f_3) n_{ph}^U \quad (10)$$

and,

$$D(E_1^{(2)} \to E_2^{(1)}) = A f_3 (1 - f_2)(n_{ph}^D + 1). \quad (11)$$

Here, $n_{ph}^U$ and $n_{ph}^D$ are the B-E distributions corresponding to photons having energies $(E_1^{(1)} - E_2^{(1)})$ and $(E_1^{(2)} - E_2^{(1)})$ respectively. We equate the up and down transitions, i.e., (10) and (11) to obtain

$$(\mu_1^{(2)} - \mu_2^{(1)}) = (E_1^{(2)} - E_2^{(1)}) - \left(\frac{T_D}{T_S}\right)(E_1^{(1)} - E_2^{(1)}). \quad (12)$$

The corresponding efficiency is,

$$\eta_{EX} = \frac{(\mu_1^{(2)} - \mu_2^{(1)})}{(E_1^{(1)} - E_2^{(1)})} = \left(\frac{(E_1^{(2)} - E_2^{(1)})}{(E_1^{(1)} - E_2^{(1)})}\right) - \left(\frac{T_D}{T_S}\right) \quad (13)$$

$$< \left(1 - \frac{T_D}{T_S}\right).$$

Note that $(E_1^{(2)} - E_2^{(1)}) < (E_1^{(1)} - E_2^{(1)})$. Thus, from (13) we find $\eta_{EX} < \eta_{Carnot}$ and the energy required to dissociate the exciton (step 2) can be viewed as thermalization loss.

To confirm that Eq. (13) is consistent with Eq. (9), recall that for a pair of donor and acceptor atoms,

$$\eta_{EX} = \frac{(N_1 + N_2)(\mu_1^{(2)} - \mu_2^{(1)})}{N_1(E_1^{(1)} - E_2^{(1)}) + N_2(E_1^{(2)} - E_2^{(2)})}.$$

In this particular case $N_2 = 0$ as there is no absorption in material B, i.e., no photon absorption involving $(E_1^{(2)} - E_2^{(2)})$. Thus we find,

$$\eta_{EX} = \frac{(\mu_1^{(2)} - \mu_2^{(1)})}{(E_1^{(1)} - E_2^{(1)})}.$$

which is exactly the term following the first equal sign in Eq. (16).

*Reducing thermalization loss*

Thermalization loss involves energy exchanged to the environment as electrons hop from one molecule to the next. Several schemes have been suggested to reduce the loss.

One approach is based on the idea of generation of multiple excitons (MEG) [16], [17]. In this scheme, the excess energy of an electron jumping from an atom to the next is not lost to phonons, but transferred to the atom acceptor atom itself, so as to thermally generate a new electron-hole pair. The atoms in level 1 is now multiply excited – first by sunlight and then by the energy of its neighbors. Thermalization loss is reduced and the efficiency approaches the Carnot limit.

Another approach to reduce thermalization loss is based on hybrid photovoltaic-thermal (PV/T) system. In this scheme, circulating fluid collects the waste heat generated by the PV module and uses the heated fluid to run an engine. (This approach should be distinguished from thermal PV or TPV). Such integrated system returns the efficiency towards the Carnot limit for systems containing multiple atoms with different bandgaps.

B. *Angular anisotropy*

In the above calculation, multiply scattered, *diffused* sunlight was used to illuminate the PV cell. Remarkably a solar cell illuminated directly by the sun has lower efficiency, as follows:

The sun is approximately $150 \times 10^6$ kilometers away, therefore it appears a small disk in the sky. Although it radiates in $4\pi$ steradians, only a fraction of this radiation, with $\theta_S = 6 \times 10^{-5}$, is incident on the earth, see Fig. 5(a). The angle is so small that the rays of sunlight can be considered parallel (and hence the shadow behind an object). On the other hand, when the photons absorbed by the atoms are re-emitted, they are radiated in all directions, i.e. radiation angle $\theta_D \sim 4\pi$ steradians. Therefore, Eq. (5) must be rewritten as,

$$\theta_D f_1(1-f_2)(n_{ph,S}+1) = \theta_S f_2(1-f_2)n_{ph,S} \quad (14)$$

$$\text{or, } -\ln\left(\frac{\theta_D}{\theta_S}\right) + \left(\frac{E_2-\mu_2}{k_B T_D}\right) + \left(\frac{E_1-E_2}{k_B T_S}\right)$$

$$= \left(\frac{E_1-\mu_1}{k_B T_D}\right)$$

$$\therefore \eta = \frac{\mu_1-\mu_2}{E_1-E_2} = \left(1-\frac{T_D}{T_S}\right) - \frac{k_B T_D}{E_1-E_2}\ln\left(\frac{\theta_D}{\theta_S}\right). \quad (15)$$

This is a remarkable formula, which says that the efficiency of a photon engine working with direct sunlight is always less than that of an engine operating in diffused light. To estimate the difference, recall that $T_S = 6000K$, $T_D = 300K$. Thus,

$$qV = \left(1-\frac{T_D}{T_S}\right)E_g - k_B T_D \ln\left(\frac{\theta_D}{\theta_S}\right)$$

$$= 0.95 \times E_g - 0.31 \quad (16)$$

Here $E_g = E_1 - E_2$. In other words, almost 30% of the open circuit voltage is lost for typical bandgap of solar cells (1-1.5eV) because of mismatch of the solar angle. For 3D solar cells, the constants are slightly different [18][19]. The loss due to angular anisotropy is partially compensated by contribution from 3D photonic density of states. Thus the open circuit voltage (3D solar cells) can be approximately represented as follows,

$$qV = 0.95 \times E_g - 0.22 \quad (17)$$

Remarkably, the best solar cells produced to date all follow Eq. (17), as shown in Fig. 5(b).

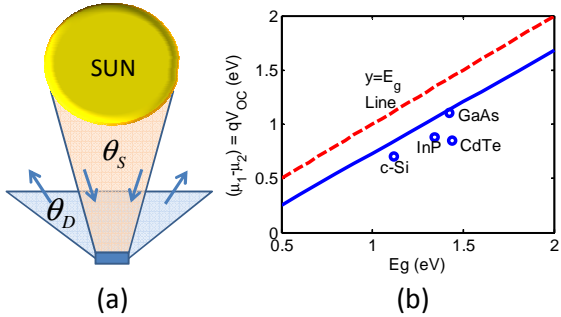

(a)      (b)

**Figure 5: (a) Angle mismatch between the sun and the solar cell. (b) The open-circuit voltage limit of a PV as a function of bandgap. The experimental results (circles) are taken from [8].**

*Diffused vs. direct sunlight*

What is the difference between diffused vs. direct sunlight that can change the PV efficiency so radically? An atom has certain directivity in the radiation pattern in vacuum [20]. In the derivation above, however, we have assumed that the photons arrive and are absorbed in a narrow angle ($\theta_S$), while they reradiate in a broader angle of $4\pi$. This can only happen if the phase of the atoms excited to level 1 are subsequently randomized by the collision among the atoms, so that the atoms eventually re-emit with random angles. The entropy gain of a system is $kT_D \ln(\text{states}_{final}/\text{states}_{initial})$ and since $(\text{states}_{final}) \sim 4\pi$ and $(\text{states}_{initial}) \sim \theta_S$, we see that the extra loss term can be viewed as an irreversible entropy gain due to angular mismatch between incident and reradiated photons. A complex derivation of this entropy loss exists [18], but the use of two-level PV makes the physical interpretation intuitive and transparent.

A classical derivation of the entropy generated produces the same results:

$$\theta_D f_1(1-f_2)(n_{ph,S}+1) = \theta_S f_2(1-f_2)n_{ph,S} \quad (18)$$

$$\therefore \frac{E_2-\mu_2}{k_B T_D} + \frac{E_1-E_2}{k_B T_S} = \frac{E_1-\mu_1}{k_B T_D} + k_B \ln\left(\frac{\theta_S}{\theta_D}\right)$$

$$\text{or, } \frac{-Q_{22}}{T_D} + \frac{Q_{12}}{T_S} = \frac{Q_{11}}{T_D} + k_B \ln\left(\frac{\theta_S}{\theta_D}\right)$$

$$\text{or, } -S_D^{(2)} + S_{sun} = S_D^{(1)} + S_{angle}$$

$$\text{or, } S_{sun} = \left(S_D^{(1)} + S_D^{(2)}\right) + S_{angle}$$

$$\text{or, } S_{in} = (S_{out}) + S_{angle} \quad (19)$$

Clearly, the angle anisotropy makes the system irreversible.

What does it mean to 'lose energy' due to angle anisotropy? Let us say that a number of photons enter the solar cell at normal incidence. The photons are absorbed and atoms are excited. The atoms then go through a momentum scattering process and subsequently, they emit at random angles. Individually, the photons have the same energy on emission as they did on absorption and there should be no loss of energy! However, we should recognize that it will take energy to create collimated photons (similar to *incident* sunlight) from random photons emitted by the cell. This energy has been irreversibly lost in the process of momentum scattering or angle randomization.

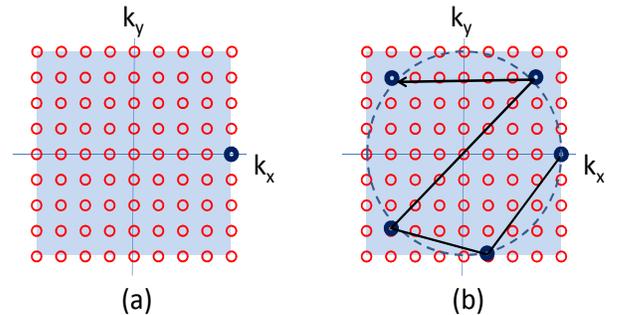

(a)      (b)

**Figure 6: (a) A single state occupied by a photon normally (approx.) incident from the sun. (b) Momentum scattering of the photon inside the solar cell.**

*Recovery of entropy loss*

The efficiency of the solar cell will improve if we can reduce the entropy loss due to angle mismatch. We can either make the absorption angle larger, or the emission angle narrower– and both approaches are in practical use today.

*Mirrors.* Solar cells often use mirrors in the back surface, which reflects light and reduces the emission angle from $4\pi$ to $2\pi$. Inserting this new angle in Eq. (15), we find that the open circuit voltage increases by, $(kT_D/q) \times \ln(2)$ or 17mV at room temperature. This leads to slight improvement in efficiency.

*Solar concentrator.* If the atoms are placed in a small sphere at the foci of a concentric hemisphere, the atoms will be illuminated from all sides with $(2\pi/\theta_S) \sim 10^5$ suns, see Fig. 7(a). The incident angle is now $2\pi$, matching exactly the angle of the radiated photons. In this case, angular anisotropy term disappears and $V_{OC}$ once again reaches the values corresponding to the Carnot limit. Therefore, the essence of the concentrator solar cells lies in countering the angle entropy generated in typical solar cell illuminated by direct sunlight.

*Narrow emission angle.* It might be possible to create a set of optical structures, so that illumination and emission are possible only with a narrow solid angle, as has been suggested in Ref. [21]. Depending on the narrowness of the angle, the efficiency should approach the Carnot efficiency for a collection of two-level atoms.

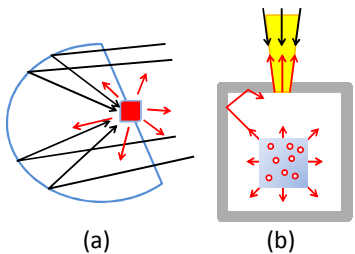

**Figure 7: (a) Angle broadening of incident photons using a solar concentrator. (b) A scheme limiting the emission angle of the PV system.**

### C. Below-Bandagp loss

Traditionally, the Shockly-Queisser efficiency of a solar cell is defined by the ratio of energy converted to electricity to total incident energy from the sun. If a photon with energy below (or above) the bandgap passes right through the atoms – never interacting with the atoms themselves – the solar cells will still be held responsible for not being able to convert it. This below-bandgap loss is really not a loss at all, because the photons still carry the memory of the sun and ability to do work. The definition presumes that the transmitted energy will be irretrievably lost, and therefore, should be rightfully chalked up as a loss mechanism.

*Recovery of below-bandgap loss*

Consider, for example, that a quasi-transparent PV has been integrated with the structure of a greenhouse. The below-bandgap photons that escapes through the solar cells can still be used to drive the photosynthesis of the plants. It is an interesting example of a 'tandem cell' for high-efficiency energy conversion. And the combined efficiency of the two system projects towards Carnot efficiency. If the PV/T absorber is opaque to below bandgap transmission, a fraction of the below-bandgap energy can also be retrieved.

The most interesting scheme to utilize the below bandgap loss involves Thermal photovoltaic (or TPV) [22]. Here, the first layer absorbs sunlight directly to heat a fluid and re-emits at a lower energy. The second selective emitter layer transmits photons that are easily absorbed by the PV layer in the bottom, but reflects to the absorber the below and above bandgap photons that have previously been lost to below-bandgap transmission and above bandgap thermalization. These 'return-to-the-sender' photons keep the top absorber layer hot and allows better conversion efficiency for the PV layer at the bottom.

### IV. SUMMARY

An idealized two-level solar cell working in diffused light is shown to achieve the thermodynamic Carnot efficiency of ~95%. In practice, however, three loss mechanisms reduce the efficiency of PV system far below the Carnot limit. The thermalization loss involves asymmetry in energy of the absorption and emission – photons are absorbed in broad-band, but emitted only in narrow band, with the rest of energy lost to phonons. Hybrid PV/T or MEG systems that recycles the waste heat improves efficiency. The second source of loss involves angle mismatch between direct illumination and emission in random angle, the so called angle entropy loss. This loss can be reduced either by reducing the emission angle by mirrors or waveguides, or increasing the incident angle by solar concentrators. Finally, the 'accounting' or below-bandgap loss can be improved by tandem cells or the TPV approach. Considerations of these loss-mechanisms – within the context of a simple two-level PV system -- collectively explain the efficiency degradation from the Carnot limit to the widely known Shockley-Queisser limit.

### APPENDIX

#### A. Two-level system illuminated by LEDs

There is an interesting corollary to the derivation of (8) presented in Sec. II.B. Consider that the two-level system is being illuminated in 3D by LEDs, rather than by the diffused sunlight (see Fig. A.1). Because of the isotropic incident light, the absorption angle and the emission angle are equal (both are $4\pi$ Steradian). Thus there will be no angle anisotropy loss. Now $(\mu_1-\mu_2)_{LED} \equiv qV_{LED} > 0$, where the LED is forward biased by $V_{LED}$. The photons emitted from the LED have nonzero chemical potential, i.e., $(\mu_1-\mu_2)_{LED} = \Delta\mu_{LED} > 0$, reflecting that fact that this source does not emit any photon with energy below the bandgap. The physical meaning of non-zero $\Delta\mu$ for nonequilibrium light sources has been discussed in Ref. [12].

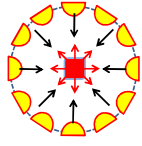

**Figure A.1: A PV system illuminated by 3D surround LED.**

A recalculation of (5) with (1) and (4) using non-zero $\Delta\mu$ produces

$$V_{oc} \equiv (\mu_1-\mu_2)_D$$
$$= (E_1-E_2)_D\left[1-\frac{T_D}{T_{LED}}\right] + (\mu_1-\mu_2)_{LED} \times \frac{T_D}{T_{LED}}. \quad (20)$$

And the efficiency is,

$$\eta = \left[1-\frac{T_D}{T_{LED}}\right] + \frac{T_D}{T_{LED}}\frac{(\mu_1-\mu_2)_{LED}}{(E_1-E_2)_D}. \quad (21)$$

Here $0 < (\mu_1-\mu_2)_{LED} < (E_1-E_2)_{LED}$ and $(E_1-E_2)_{LED} \geq (E_1-E_2)_D$. The second inequality follows from the requirement that the LED must emit photons at energies that the atoms can absorb. Moreover, one assumes that the emission from the device does not affect the Fermi-level from the source LED. Under these conditions, we find $1 > \eta > (1-T_D/T_S)$ – that the system exceeds the Carnot efficiency.

This intriguing result can be interpreted as follows: Carnot engine is assumed to operate between two reservoirs defined by temperature T and chemical potential $\mu$. If the reservoir itself is not in thermodynamic equilibrium, with splitting of its own electro-chemical potential – as is the case for LEDs – exceeding the Carnot efficiency is not impossible.

Of course, when accounted for the electrical energy necessary for the LED to work – the overall efficiency return to the Carnot limit. In all fairness, we also do not account for the nuclear reaction in the sun in our calculation of energy balance. In that strict sense, even a solar illuminated two-level system may exceed the Carnot limit – although the margin of gain is likely to be infinitesimal.

Finally, one could say that the photons emitted from the LED has a higher effective temperature $T^*_{LED}(>T_{LED})$ and write,

$$\eta = \left[1-\frac{T_D}{T_S}\right] + \frac{T_D}{T_S}\frac{(\mu_1-\mu_2)_{LED}}{(E_1-E_2)_D} \equiv \left[1-\frac{T_D}{T^*_{LED}}\right], \quad (22)$$

suggesting that Carnot limit is preserved with redefined temperature. However, Fig. 1c shows that this is not quite correct because the Bose-Einstein distribution, $n_{ph}$ for different chemical potentials ($\mu$) cannot be made equal by simply modifying the temperatures, i.e., $n_{ph}(\mu_1,T_1) \neq n_{ph}(0,T^*_1)$.

ACKNOWLEDGMENT

We gratefully acknowledge discussion with Profs. J. Gray, M. Lundstrom, and P. Bermel. This material is based upon work supported as part of the Center for Re-Defining Photovoltaic Efficiency Through Molecule Scale Control, an Energy Frontier Research Center funded by the U.S. Department of Energy, Office of Science, Office of Basic Energy Sciences under Award Number DE-SC0001085.The computational resources for this work were provided by the Network of Computational Nanotechnology under NSF Award EEC-0228390.